\documentclass[11pt]{article}

\usepackage{osid}
\usepackage{latexsym}
\usepackage{amsfonts}
\usepackage{graphicx}
\usepackage{amssymb}
\usepackage{amsmath}

\newcommand{\bra}[1]{\langle #1 |}
\newcommand{\ket}[1]{| #1 \rangle}

\newcommand{\matx}[1]{| #1 \rangle \langle #1 |}

\title{Effect of entanglement on geometric phase for multi-qubit states}
\author{Mark S. Williamson \\{\footnotesize\it School of Physics and Astronomy, University of Leeds,
 Leeds LS2 9JT, UK.}%\\
 \\[2ex]
        Vlatko Vedral%\thanks{Supported by ...}
                     \\{\footnotesize\it
                     School of Physics and Astronomy, University of Leeds, Leeds LS2 9JT, UK, \\
                     Centre for Quantum Technologies, National University of Singapore, 3 Science Drive 2,
                     Singapore 117543,\\
                     Department of Physics, National University of Singapore, 2 Science Drive 3,
                     Singapore 117543.\\
                     E-mail: mark.williamson@quantuminfo.org} }

\begin{document}
\maketitle

\begin{abstract}
When a multi-qubit state evolves under local unitaries it may obtain a geometric phase, a feature dependent on
the geometry of the state's projective Hilbert space. A correction term to this geometric phase in addition
to the local subsystem phases may appear from correlations between the subsystems.
We find this correction term can be characterized completely either by the entanglement or completely
by the classical correlations for several classes of entangled state. States belonging to the former set are W states and
their mixtures, while members of the latter set are cluster states, GHZ states and two classes of bound
entangled state. We probe the structures of these states more finely using local invariants and suggest the
cause of the entanglement correction is a gauge field like $SL(2,\mathbb{C})$ invariant recently introduced
named twist.
\end{abstract}

\section{Introduction}

The phenomenon of quantum entanglement has received widespread
attention recently as researchers have recognized its importance in
quantum information theory. Beyond two qubits not much is known
about entanglement. Its characterization and quantification becomes
particularly hard as the number of possibilities a multi-qubit
system can be entangled grows with the number of qubits. For a
comprehensive review of entanglement see \cite{ref:Horodecki07}
and references within.

Previous workers have studied the geometric phases of entangled states, mainly restricted to two qubits in a pure state \cite{ref:Sjoqvist00a,ref:Tong03b,ref:Yi04b,ref:Basu06,ref:Sjoqvist08}.
The geometric phase is a well understood and
celebrated effect resulting from the geometry of the state's
projective Hilbert space \cite{ref:ShapereBook}. In this paper we study the effect multi-qubit
entanglement has on the geometric phase in an attempt to distill the geometric features
of entanglement. We imagine
an entangled $N$ qubit state where each of the $N$ qubits are spatially separated and are in the possession
of $N$ parties. Each party may only perform (local) unitaries, analogous to local gauge transformations,
on their own qubit. In this way the entanglement and
nonlocal properties of the state must remain fixed but it may still obtain a geometric phase dependent
on the geometry of its projective Hilbert space.
We examine the difference entanglement
makes to this phase and therefore to this geometry. Our hope is to
elucidate which of the plethora of entanglement structures possible in multi-qubit
systems characterized by locally invariant functions of the state parameters are responsible
for altering this geometry.

In particular we attempt to understand the following observation:
\emph{Quantum or classical correlations between subsystems
in a composite state modify the geometric phase under local unitary evolution}.
Stated another way the overall geometric phase of a correlated state $\Gamma$ cannot be
written as the sum of its parts, there is a correction term dubbed
the mutual geometric phase $\Delta \gamma$ in addition to the local
geometric phases obtained by the individual subsystems we label
$\gamma_n^M$. The subscript $n$ labels indexes each of the $N$
subsystems in the correlated state and the superscript $M$ for mixed
refers to the fact that in general the subsystems will be in mixed
states described by density matrices. We can write this as
\begin{equation}
\Gamma=\Delta \gamma + \sum_{n=1}^N \gamma_n^M.
\end{equation}
This is \emph{not} true of the other phase in quantum mechanics, the
dynamical phase. When we restrict to local unitary evolution, the overall
dynamical phase of the correlated state
$\Upsilon$ can always be understood as the sum of its subsystem's
dynamical phases $\upsilon_n^M$. One can verify this from the
definition of dynamical phase \cite{ref:Sjoqvist00}
\begin{equation}
\Upsilon=-i\int_0^T\hbox{tr}[\rho \mathcal{U}(t)^\dag
\dot{\mathcal{U}}(t)]dt
\end{equation}
and the local unitary condition $\mathcal{U}(t)=\bigotimes_{n=1}^N
U_n(t)$. Differentiating $\mathcal{U}(t)$ with respect to
$t$ and plugging it back into the equation for dynamical phase we
find
\begin{equation}
\Upsilon=\sum_{n=1}^N \upsilon_n^M.
\end{equation}
We have not assumed anything about the composite state $\rho$, it
can be completely general; entangled, classically correlated or
uncorrelated. Correlations of any type make no difference. This statement can
be seen to be trivial when we
regard the local unitaries to effectively model local dynamics. In
restricting the dynamics to be local we see the composite dynamical
phase can also be thought of as local. In contrast, it can also be
seen the geometric phase given by the equation \cite{ref:Sjoqvist00}
\begin{equation}\label{eq:geophase}
\Gamma=\arg\{\hbox{tr} \rho \mathcal{U}^\shortparallel(T)\}
\end{equation}
\emph{is} modified by correlations under the local unitary
condition. An uncorrelated, product state
$\rho=\bigotimes_{n=1}^N \rho_n$ can however be written
as the sum of its local, subsystem phases as one would expect.
$\mathcal{U}^\shortparallel$ is the unitary implementing parallel
transport on a given path. We will explain what this means in more
detail in section \ref{sec:paralleltransport}

Using ideas from entanglement distance measures we can determine
which correlations are responsible for this modification of the
geometric phase. Correlations are divided into the two coarsest
categories by these measures: quantum correlations
(entanglement) and classical correlations. From these ideas we
calculate three geometric phases associated to a given state (i) the
geometric phase of the composite entangled state (ii) the geometric
phase of the closest separable state, the state with only the
classical correlations between subsystems present
and (iii) the geometric phase of the uncorrelated state, that is the
composite entangled state with all correlations
removed. By comparing these phases we can see what effect the entanglement and the
classical correlations have on the geometric phase and therefore the
geometry of the projective state space. This is explained in section~\ref{sec:mutgeo}

As entanglement is defined in distance measures as being
the surplus correlation not able to be described by classical
correlations alone one would intuitively believe that $\Delta
\gamma$ can be attributed to a mixture of both entanglement and
classical correlations. We find however that the states analyzed
belong to one of two sets: the modification $\Delta \gamma$ is due
\emph{only} to entanglement \emph{or} the modification $\Delta
\gamma$ is due \emph{only} to the classical correlations. We find
that W like states and mixtures of W and \={W} states belong to the
former set while Greenberger-Horne-Zeilinger (GHZ) states, cluster
and two types of bound entangled state, those of D\"{u}r and Smolin
belong to the latter set. We also find that for pure states of two
qubits the mutual geometric phase $\Delta \gamma$ is always
accounted for by classical correlations. For entanglement to affect
the geometry of the projective state space one at least needs composite states
of three qubits or more. First we review and extend previous work \cite{ref:Williamson&Vedral07}
using these ideas in section~\ref{sec:review} and present new analysis in
section~\ref{sec:otherstates}

In an attempt to understand which features of an entangled state may
be responsible for these results we look at the local
invariants of the state. That is the things about the entangled
state that do not change under local unitaries like the
amount of entanglement for instance. It is known that there is only
one local invariant of a two pure qubit state, it characterizes the amount of
entanglement in the state. For three or more qubits the structures get
much richer. One needs five local invariants to describe an arbitrary
pure state of three qubits $a$, $b$ and $c$, only four of which have a clear meaning.
Three can be thought of as the bipartite entanglements; how
entangled $a$ is with $b$, $b$ with $c$ and $c$ with $a$. There is also the
3-tangle, how entangled $a$, $b$ and $c$ are together in a three way
correlation and lastly there is the Kempe invariant
which seems to have a more geometrical origin following some
recent work \cite{ref:Williamson&Wootters08}.
We calculate and compare these invariants for the various states in the
hope of shedding some light on the cause of two distinct results. In section
\ref{sec:statecharacteristics} we show evidence that a local invariant named twist, a function of the Kempe invariant, may be the cause of the modification of geometric phase when entanglement is responsible before finally concluding.

\section{Correlations responsible for the difference in geometric
phase}\label{sec:previous}

In this section we review the core of the analysis, how we characterize
$\Delta \gamma$. These calculation are illustrated in detail
for GHZ and W state, two
inequivalent forms of entanglement under stochastic local operations and
classical communication (SLOCC) \cite{ref:Dur00}, structures first appearing in pure states of three qubits. First we will review mixed state parallel transport conditions from which one may obtain geometric phases.

\subsection{Mixed state parallel transport conditions and geometric phases}\label{sec:paralleltransport}
If at each neighboring point along a state's path it is in phase with
itself any global phase obtained will be purely geometrical in origin. Moving a state around in this manner is known as parallel transport. Non-trivial
parallel transport around a closed loop indicates the space over which the parallel
transport is taking place has some curvature. In the case of
pure quantum states, parallel transport effectively means no dynamical phase is obtained over
the path taken. For clarity, once a pure state $\ket{\psi}$
completes a closed path parameterized by $t$ in projective Hilbert space by the unitary $U(t)$ it will have
picked up a global
phase $e^{i(\gamma+\upsilon)}\ket{\psi}$. If the state is parallel transported
over this path, the dynamical phase $\upsilon=0$ and one is left only with the
geometric phase $\gamma$. Mathematically the condition for parallel
transport can be written $\bra{\psi}U(t)^\dag \dot{U}(t)\ket{\psi}=0$. We write
the unitary that fulfills these parallel transport
conditions $U^\shortparallel$(t).

One can also define parallel transport conditions
for mixed states in which case there are multiple choices. We
work with the stronger parallel transport condition of \cite{ref:Sjoqvist00}.
These conditions are known to
produce a geometric phase that is a property of the mixed state alone
\cite{ref:Ericsson03b} and require each eigenvector $\ket{\phi_i}$ of the mixed
state $\rho=\sum_i \lambda_i \matx{\phi_i}$ to be parallel transported i.e.
\begin{equation}\label{eq:paralleltransport}
\bra{\phi_i}U(t)^\dag \dot{U}(t)\ket{\phi_i}=0, \forall i.
\end{equation}
Once we have parallel
transported a state we know its total overall phase will be the geometric
phase. In this case we can use eq.~(\ref{eq:geophase}), the equation for the total phase,
to calculate its geometric
phase. Eq.~(\ref{eq:geophase}) will be used to calculate geometric phases in all the following
analysis. Incidentally this formula is valid for all paths, not just closed,
cyclic evolutions but in this paper we will parallel transport each subsystem
of the entangled state cyclically.

As an example how one might calculate a specific geometric phase associated
to a particular Hamiltonian and path imagine a qubit in the state
$\ket{0}$ precessing around an axis at angle $\theta$ to the $z$ axis in the $x-z$ plane on the
Bloch sphere at frequency $\omega$. The Hamiltonian corresponding to this
precession in the $\ket{0}$, $\ket{1}$ basis is
\begin{equation}\label{eq:hamiltonian}
H=\frac{\omega}{2}\left(
                    \begin{array}{cc}
                      \cos \theta & \sin \theta \\
                      \sin \theta & -\cos\theta \\
                    \end{array}
                  \right).
\end{equation}
The unitary is then $U(t)=e^{-iHt}$ however this is not the unitary that implements
parallel transport. To find this we need to consider which set of unitaries trace
the same path in the projective Hilbert space for a given mixed state
$\rho=\sum_i^d \lambda_i \matx{\phi_i}$. It is the set
\begin{equation}
\tilde{U}(t)=U(t)V(t).
\end{equation}
$V(t)$ is a unitary that commutes with $\rho$ i.e. $[V(t),\rho]=0$. One can verify
$\tilde{U}(t)\rho \tilde{U}(t)^\dag = U(t)\rho U(t)^\dag$ i.e. they both trace the
same path. In the case of a mixed state with non-degenerate
eigenvalues the most general $V(t)$ is
\begin{equation}\label{eq:gauge}
V(t)=\sum_i^d e^{i\varphi_i(t)}\matx{\phi_i}.
\end{equation}
This gauge transformation belongs to the group $U(1)^d$ written in the eigenbasis of the density matrix.
In the case of a degenerate density matrix with degeneracy $m$ the symmetry group of the gauge transformation
that results in the same path is enhanced to $U(m)\times U(1)^{d-m}$. As an example imagine the three level
density matrix $\rho=\matx{0}+\matx{1}+2\matx{2}$. This state traces the same path not only with $V(t)$ given
by eq.~(\ref{eq:gauge}) ($U(1)\times U(1)\times U(1)$) but also under the group $U(2)\times U(1)$. The resulting
geometric phase factor will then be non-abelian, see \cite{ref:Singh03} for further details. Ultimately the
symmetries are determined by the physics of the problem and in this study where we imagine each qubit subsystem
to be parallel transported, spatially separated from the others, we have at most $U(2)$ gauge symmetries when
the subsystems are maximally mixed. Even when these cases occur we will restrict to symmetries given by
eq.~(\ref{eq:gauge}). In other words we will calculate geometrical, gauge invariant $(U(1)\times U(1))^N$
structures of the projective Hilbert space, $N$ being the number of qubits in the state.

\begin{figure}
\begin{center}
\includegraphics[width=8.6cm]{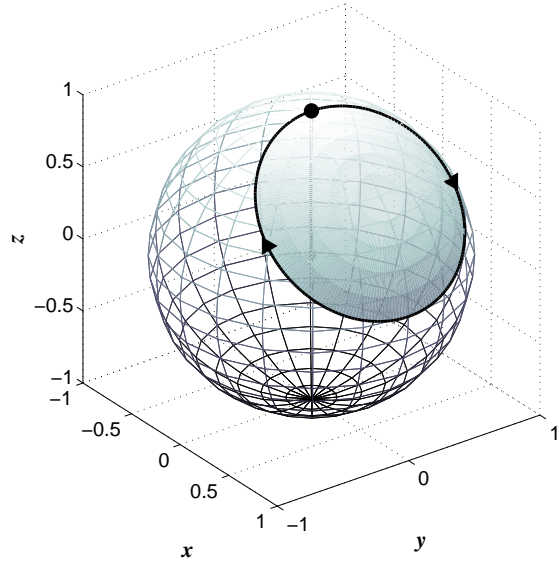}
\end{center}
\caption{Cyclic evolution of the state $\ket{0}$ for the Hamiltonian in eq.~(\ref{eq:hamiltonian}) on
the Bloch Sphere. The state $\ket{0}$ is represented by the North pole and the black line traces the
path the state makes during its evolution. The solid angle enclosed by the path $\ket{0}$ sweeps out
illustrated by the shaded area, $\Omega$, is proportional to the geometric phase, $\gamma$.
On the unit Bloch sphere $\gamma=\Omega/2$. For this example the angle of the precession axis
from the $z$ axis is $\theta=\pi/4$.}\label{fig:BlochSphere}
\end{figure}

Restricting to this abelian case, we need to find the $\tilde{U}(t)$ that implements parallel
transport by solving for $\varphi_i(t)$ using the parallel transport conditions. This results
in the parallel transporter being
\begin{equation}
U^\shortparallel(t)=U(t)\sum_i^d e^{-\int_0^t \bra{\phi_i}U(t')^\dag \dot{U}(t')\ket{\phi_i}dt'}\matx{\phi_i}.
\end{equation}
One can verify that this choice of $U(t)$ results in a $V(t)$ invariant geometric phase. For the specific
Hamiltonian and
state $\ket{0}$ considered in this example the geometric phase after a cyclic evolution, $T=2\pi/\omega$ is
\begin{eqnarray}
&\gamma=\arg \bra{0}U^\shortparallel(T) \ket{0}=
\arg\{\bra{0}U(T)\ket{0}e^{-\int_0^T \bra{0}U^\dag(t)\dot{U}(t)\ket{0}dt}\} \nonumber\\
&=-\pi(1-\cos\theta).
\end{eqnarray}
The geometric phase is proportional to the area enclosed by the path and in figure~\ref{fig:BlochSphere}
we have illustrated this example. In the work that follows we will work more generally without referring
to a specific Hamiltonian, making the identifications
\begin{eqnarray}\label{eq:purestateidentities}
&\bra{0_n}U_n^\shortparallel(T) \ket{0_n}=e^{i\gamma_n}\\
&\bra{1_n}U_n^\shortparallel(T) \ket{1_n}=e^{-i\gamma_n}
\end{eqnarray}
where $n$ refers to the subsystem. The $\pm \gamma_n$ are the geometric phases the pure states $\ket{0}$
and $\ket{1}$ obtain over the arbitrary cyclic evolution $U_n(T)$. Alternatively one can
view $\gamma_n$ as half the solid angle enclosed by the path of $\ket{0_n}$. One can verify that the state
$\ket{1}$ does indeed pick up an equal and opposite geometric phase to $\ket{0}$. By looking at
figure~\ref{fig:BlochSphere} one can see that this must be the case. The unitary preserves the scalar product
between states and since $\ket{0}$ and $\ket{1}$ are orthogonal, $\ket{1}$ must trace the same path as $\ket{0}$
on the opposite side of the Bloch sphere but in the anti-clockwise rather than clockwise direction giving
the minus sign. One notes that this type of structure described by just one parameter, $\gamma$, will not be present for subsystems with more than two levels.

\subsection{Determining which correlations are responsible for the mutual geometric phase}\label{sec:mutgeo}

Our next step is to determine which correlations are responsible for $\Delta\gamma$ in the expression
\begin{equation}
\Gamma=\Delta \gamma +\sum_{n=1}^N \gamma_n^M.
\end{equation}
We can do this by calculating three geometric phases. (i) The geometric phase of the entangled state $\Gamma$
(ii) the geometric phase of the closest separable state (just classical correlations) $\Gamma_{sep}$ and
(iii) the geometric phase of the
uncorrelated state, the subsystem states tensored together $\sum \gamma_n^M$. By splitting
$\Delta \gamma$ into entanglement $\Delta \gamma_q$ and classical correlation $\Delta \gamma_c$ contributions
so that $\Delta \gamma=\Delta \gamma_q + \Delta \gamma_c$ we see which correlations contribute to
$\Delta \gamma$. Defined in this way the difference entanglement makes to the geometric phase is
\begin{equation}\label{eq:deltagammaq}
\Delta \gamma_q=\Gamma-\Gamma_{sep}.
\end{equation}
Likewise we can see the difference classical correlations make to the geometric phase using
\begin{equation}\label{eq:deltagammac}
\Delta \gamma_c=\Gamma_{sep}-\sum_{n=1}^N \gamma_n^M.
\end{equation}
In other words the difference between the geometric phases of the maximally classically correlated state and the
uncorrelated, product state obtained by tracing each of the subsystems out of our entangled state.

\begin{figure}
\begin{center}
\includegraphics[width=7cm]{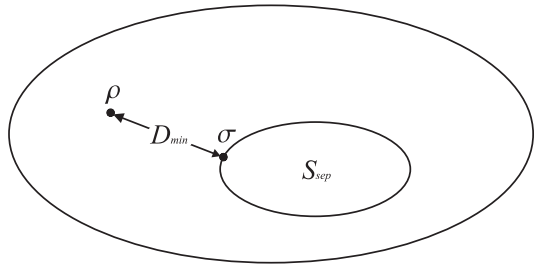}
\end{center}
\caption{The boundary of the set of all states is the outermost one. The inner set is the set of all
separable states. The two points $\rho$ and $\sigma$ are the entangled state and the closest separable
state respectively. The amount of entanglement in $\rho$ is given by the minimum distance $D_{min}$
between itself and the closest point on the set of separable states, $\sigma$. $D$ can be any measure,
for the relative entropy of entanglement the measure is the relative entropy and $D_{min}=E_R$.}
\label{fig:entdistancemeasures}
\end{figure}

How do we find the closest separable state, $\sigma$? This
is the state from the set of all separable states, $S_{sep}$, that minimizes the relative entropy between it and
the entangled state, $\rho$. The relative entropy of entanglement, $E_R$, is defined as this minimum
\cite{ref:Vedral97}
\begin{equation}
E_R=\min_{\sigma \in S_{sep}} \hbox{tr} \left(\rho \log \rho-\rho\log\sigma\right).
\end{equation}
It is probably the most
fundamental of a family of entanglement measures called entanglement distance measures. The idea of these
measures is that entanglement is defined as the minimum surplus correlation that cannot be accounted for
just by classical correlations. The state $\sigma$ replicates as much of the correlation in $\rho$ as possible
while only being allowed to be separable. A schematic of this idea is shown in
figure~\ref{fig:entdistancemeasures}. These measures are attractive as they apply to systems of any dimension
(any number of qu$d$its). The information they provide however is quite coarse, telling you only how much
entanglement is in a given state and not the character of the entanglement (they will not tell you
whether the entanglement is bipartite/tripartite etc). Distance measures are not easy to calculate either. The
hard part of the problem is finding the state that minimizes the given distance measure. This is also the factor
that constrains the work in this paper to the classes of entangled states with known $\sigma$.

\subsection{Geometric phases for GHZ and W states}\label{sec:review}

We now calculate geometric phases and characterize $\Delta\gamma$ for the GHZ and W states.
We write the $N$ qubit GHZ state as
\begin{equation}
\ket{GHZ}=\alpha\ket{0}^{\otimes N}+\beta\ket{1}^{\otimes N}.
\end{equation}
$\alpha$ and $\beta$ can be made real without loss of generality by making local transformations. GHZ states only
have entanglement at the full $N$ qubit level. Once a single qubit is lost the state is separable. W states on the
other hand remain entangled down to the last pair of qubits. Our W state is
more general than what is usually referred to as a W state in the literature.
These states will be written $\ket{N,k}$
where $N$ refers to the number of qubits and $k$ refers to how many are in the state $\ket{1}$. The
states are an equal symmetric superposition of all possible distinct permutations.
\begin{equation}
\ket{N,k}=\frac{1}{\sqrt{{\footnotesize \left(
                           \begin{array}{c}
                             N \\
                             k \\
                           \end{array}
                         \right)}}}\hat{S}\ket{\underbrace{000}_{N-k}....\underbrace{111}_k},
\end{equation}
$\hat{S}$ is the complete symmetrization operator. As an example the familiar W state is written
$\ket{3,1}=\frac{1}{\sqrt{3}}(\ket{100}+\ket{010}+\ket{001})$.

First we compute the composite geometric phases for these states
using eq.~(\ref{eq:geophase}) under
the conditions that the subsystems are parallel transported according to the mixed state conditions
eq.~(\ref{eq:paralleltransport}) and the evolution of the subsystems is cyclic meaning each subsystem comes
back to the same ray for example $\ket{0}=U^\shortparallel(T)\ket{0}$ up to a phase factor. We will parallel
transport each of the subsystems locally so
\begin{equation}
\mathcal{U}^\shortparallel(t)=\bigotimes_{n=1}^N U_n^\shortparallel(t)
\end{equation}
where $U_n(t) \in SU(2)$. For the GHZ state we have
\begin{equation}
\Gamma_{GHZ}=\arg\left\{ \alpha^2 \prod_{n=1}^N \bra{0}U_n^\shortparallel(T)\ket{0} +
\beta^2 \prod_{n=1}^N \bra{1}U_n^\shortparallel(T)\ket{1}\right\}\nonumber
\end{equation}
\begin{equation}
=\arg\left\{ \alpha^2 e^{i\sum_{n=1}^N \gamma_n} + \beta^2 e^{-i\sum_{n=1}^N \gamma_n}\right\},
\end{equation}
and for the W state
\begin{equation}\label{eq:Wstategeophase}
\Gamma_W=\arg\left\{\frac{1}{\footnotesize \left(
                               \begin{array}{c}
                                 N \\
                                 k \\
                               \end{array}
                             \right)} \sum_{m=1}^{\tiny \left(
                               \begin{array}{c}
                                 N \\
                                 k \\
                               \end{array}
                             \right)} e^{i\sum_{n=1}^N A_{mn} \gamma_n}\right\}.
\end{equation}
In the equation for $\Gamma_W$ we have introduced the ${\tiny \left(
                               \begin{array}{c}
                                 N \\
                                 k \\
                               \end{array}
                             \right)}$ by $N$ matrix $A$ to capture the sign of $\gamma_n$. Each
row has $N-k$ elements that are $1$ and $k$ elements being $-1$. Each row is a distinct permutation
of the elements of this first row.

Next we calculate the local, subsystem geometric phases, $\gamma_n^M$, of the two states. To do
this we first find the subsystem states, $\rho_n$, by tracing out all but the subsystem from $\ket{GHZ}$ or
$\ket{N,k}$ we are interested in. Because of the permutation symmetry all the $N$ subsystems have the same
state.
\begin{eqnarray}
\rho_n^{GHZ}=\alpha^2\matx{0}+\beta^2\matx{1},\\
\rho_n^W=\frac{N-k}{N}\matx{0}+\frac{k}{N}\matx{1}.
\end{eqnarray}
From these subsystem states we can calculate the local subsystem phases of the uncorrelated ($uc$) state
$\rho^{uc}=\bigotimes_{n=1}^N \rho_n$, the local geometric phases $\sum \gamma_n^M$. For the GHZ state we have
\begin{equation}
\left(\sum_{n=1}^N \gamma_n^M\right)^{GHZ}=
\sum_{n=1}^N \arg\left\{\alpha^2 e^{i\gamma_n}+\beta^2 e^{-i\gamma_n}\right\}\nonumber
\end{equation}
\begin{equation}
=\arg\left\{\prod_{n=1}^N \alpha^2 e^{i\gamma_n}+\beta^2 e^{-i\gamma_n}\right\},
\end{equation}
and the W state
\begin{align}
&\left(\sum_{n=1}^N \gamma_n^M\right)^{W}=\arg\left\{\prod_{n=1}^N \frac{N-k}{N} e^{i\gamma_n} +
\frac{k}{N}e^{-i\gamma_n}\right\} \nonumber\\
%\end{equation}
%\begin{equation}
&=\arg\left\{\frac{1}{N^N}\sum_{l=0}^N (N-k)^{N-l} k^l \sum_{m=1}^{\tiny \left(
                               \begin{array}{c}
                                 N \\
                                 l \\
                               \end{array}
                             \right)} e^{i \sum_{n=1}^N A^l_{mn}\gamma_n}\right\}.
\end{align}
In the last equation we have constructed another matrix $A^l$ similar to our last $A$ matrix
in eq~(\ref{eq:Wstategeophase}). The difference is that the rows of $A^l$ have $N-l$ elements with the value $1$
and $l$ elements with the value $-1$. Again, the other ${\tiny \left(\begin{array}{c}N \\l \\ \end{array}\right)}-1$
rows are the distinct permutations of the first row.

We also calculate the geometric phase of the other relevant state, the closest separable state, $\sigma$.
For the GHZ and W state these closest separable states are known. They are given by
\cite{ref:Vedral&Plenio98,ref:Wei04}
\begin{equation}
\sigma^{GHZ}=\alpha^2 \matx{0}^{\otimes N} + \beta^2 \matx{1}^{\otimes N},
\end{equation}
\begin{equation}
\sigma^{W}=\frac{1}{N^N}\sum_{l=0}^N  \left(
                                      \begin{array}{c}
                                        N \\
                                        l \\
                                      \end{array}
                                    \right)
(N-k)^{N-l} k^l \matx{N,l}.
\end{equation}
The geometric phases for these states are
\begin{equation}
\Gamma_{sep}^{GHZ}=\arg\left\{ \alpha^2 e^{i\sum_{n=1}^N \gamma_n} + \beta^2 e^{-i\sum_{n=1}^N \gamma_n}\right\},
\end{equation}
\begin{equation}
\Gamma_{sep}^W=\arg\left\{\frac{1}{N^N}\sum_{l=0}^N (N-k)^{N-l} k^l \sum_{m=1}^{\tiny \left(
                               \begin{array}{c}
                                 N \\
                                 l \\
                               \end{array}
                             \right)} e^{i \sum_{n=1}^N A^l_{mn}\gamma_n}\right\}.
\end{equation}

We now have all the ingredients necessary to characterize $\Delta \gamma$. For GHZ states one finds
\begin{equation}
\Gamma^{GHZ}=\Gamma_{sep}^{GHZ}
\end{equation}
so that $\Delta \gamma^{GHZ}=\Delta \gamma_c^{GHZ}$ and $\Delta \gamma_q^{GHZ}=0$. For GHZ states classical
correlations are solely responsible for the change in the geometric phase above the local phases. Since all
pure two qubit states can be cast in the form of a GHZ state by local transformations, this statement is also
true of all two qubit pure states.

For W states one finds the polar opposite
\begin{equation}
\Gamma_{sep}^W=\left(\sum_{n=1}^N \gamma_n^M \right)^W
\end{equation}
so that $\Delta \gamma^W=\Delta \gamma^W_q$ and $\Delta \gamma_c^W=0$. For W states entanglement is solely
responsible for the change in the geometric phase above the local phases.

For GHZ states when $\alpha=\beta$ and when $k=N/2$ for W states all geometric phase factors
are $1$ or $-1$, elements of $\mathbb{Z}_2$ giving phases of 0 or $\pi$. This happens because
the functions inside the argument in eq.~(\ref{eq:geophase}) become real. Incidently this occurs when $E_R$
is maximal for these states. In this paper we will term geometric phase factors in $\mathbb{Z}_2$ as trivial.

\section{Other states: Bound entangled, W mixtures and cluster
states.}\label{sec:otherstates}

The results from the last section are intriguing and also rather
mysterious. Since both classes of state contained both entanglement
and classical correlations one might have suspected that this would
have been reflected in type of correlation responsible for the
difference in the geometric phase. However we found two extreme
cases; the difference in W states was described purely by the
entanglement and for GHZ states it was described purely by the classical
correlations. In this section we investigate other entangled states
for which the closest separable states are known. The aim being to
pick out the features responsible for this result. We calculate
for the interesting classes of cluster states, the bound entangled states of D\"{u}r and
Smolin and mixtures of W states. As in the
last section we find these new classes of state can also be
categorized as having a either a geometric phase difference arising
solely from entanglement or classical correlations. We group these two sets
in the following subsections. Note that many of the states we write down are
unnormalized.

\subsection{$\Delta \gamma=\Delta \gamma_c$: State geometries described by classical correlation}
\subsubsection{Cluster states}
Cluster states first appear for four qubit spaces and form a new SLOCC class
\cite{ref:Osterloh&Siewert05}. They are interesting
because they have properties somewhere in the middle of GHZ and W states \cite{ref:Briegel&Raussendorf01},
remaining entangled until $N/2$ of the particles are traced out.

One may create a $N$ qubit cluster state, $\ket{\varphi^N}$, by taking $N$ pure qubits each in the state
$\ket{0}+\ket{1}$ and applying a controlled phase gate (CZ) between neighbors. The CZ in the $\ket{0}$,
$\ket{1}$ basis is the $4\times 4$ matrix $diag\{1,1,1,-1\}$. Here we consider linear cluster
states, states where CZs are applied between qubits $1$ and $2$, $2$ and $3$ etc. The
first five of these states (up to local unitary transforms) are given by
\begin{align}
&\ket{\varphi^2}=\ket{00}+\ket{11}\\
&\ket{\varphi^3}=\ket{000}+\ket{111}\\
&\ket{\varphi^4}=\ket{0000}+\ket{0111}+\ket{1011}+\ket{1100}\\
&\ket{\varphi^5}=\ket{00000}+\ket{00111}+\ket{11011}+\ket{11100}\\
&\ket{\varphi^6}=\ket{000000}+\ket{000111}+\ket{011011}+\ket{011100}...\nonumber\\
&                +\ket{101011}+\ket{101100}+\ket{110000}+\ket{110111}.
\end{align}
The two and three
qubit states are equivalent to Bell and GHZ states respectively while the four qubit state is distinct.
If we trace qubits out of the $N\ge4$ cluster states to obtain a 3 qubit state we find some partitions
are entangled. One can verify this using the Peres-Horodecki criterion
\cite{ref:Peres96,ref:Horodecki96} by transposing one of the qubits and checking if the resulting matrix is
no longer positive. Strangely, we find that even though the state is entangled it has no bipartite
or tripartite entanglement as defined by the 2- and 3-tangles (see section~\ref{sec:statecharacteristics}). It
is another example of an entangled mixed three qubit state having no 2- or 3-tangle in addition to those found by
\cite{ref:Lohmayer06}.

The general method for finding the closest separable states is given in \cite{ref:Hajdusek&Vedral08}. Using
this method we can construct the closest separable state to $\ket{\varphi^N}$. Here we give the closest state
to $\ket{\varphi^4}$
\begin{align}
&\sigma^4=\matx{0000}+\matx{0111}...\nonumber\\
&+\matx{1011}+\matx{1100}.
\end{align}
The other closest states may be constructed from $\matx{\varphi^N}$ simply by removing the off-diagonal terms in
the $\ket{0}$, $\ket{1}$ basis. Each of the $N$ subsystems are given by the maximally mixed state
\begin{equation}
\rho_n=\matx{0}+\matx{1}.
\end{equation}

One can verify that $\Gamma=\Gamma_{sep}$ and therefore $\Delta \gamma=\Delta \gamma_c$. Also
notice that the local geometric phase factors are always trivial. The
geometric phase factors of the entangled and closest separable states can however be complex
giving a continuum of possible phases.

\subsubsection{Smolin's unlockable bound entangled state}

In \cite{ref:Smolin01} Smolin presented a 4 qubit bound entangled state. Bound entangled meaning that
no pure state entanglement may be distilled from the state by LOCC. It is termed unlockable
because when two parties come together a Bell state may be obtained by the other two parties using only
LOCC. The state is
\begin{equation}
\rho^{Smolin}=\frac{1}{4}\sum_{i=0}^3 \matx{X_i}
\end{equation}
where $\ket{X_0}=\ket{0000}+\ket{1111}$, $\ket{X_1}=\ket{0011}+\ket{1100}$, $\ket{X_2}=\ket{0101}+\ket{1010}$ and
$\ket{X_3}=\ket{0110}+\ket{1001}$ are GHZ states. Once a qubit is removed the state is separable. The closest
separable state to $\rho^{Smolin}$ has been given by \cite{ref:Wei04a} and is obtained again by removing the
off-diagonal elements in this basis. Each
subsystem is given by the maximally mixed state and a straight forward calculation reveals that
$\Gamma=\Gamma_{sep}$. For this state all geometric phase factors are trivial.

\subsubsection{D\"{u}r's bound entangled state}
D\"{u}r found a state that demonstrated bound entanglement does not necessarily imply one can find a
local hidden variable model (LHV) describing the state \cite{ref:Dur01}. The violation of a Bell type
inequality indicates the non-existence of a LHV and the state D\"{u}r presented violated such an
inequality for $N\ge 8$. It was also demonstrated that for $N\ge 4$ the following state is bound entangled
\begin{equation}
\rho^{D\ddot{u}r}=x\matx{GHZ}+\frac{1-x}{2N}\sum_{k=1}^N P_k + \bar{P}_k
\end{equation}
for $x=1/(N+1)$. Wei \emph{et al.} \cite{ref:Wei04a} show that this state is bound entangled for $0<x\le 1/(N+1)$
and entangled for $x>1/(N+1)$. $\ket{GHZ}=(\ket{0}^{\otimes N}+\ket{1}^{\otimes N})/\sqrt{2}$ and
$P_k=\ket{0}_1\ket{0}_2\ket{0}_3...\ket{1}_k...\ket{0}_N$ i.e. a projector composed of $\ket{0}$s but with
$\ket{1}$ in the $k$th qubit position. $\bar{P}_k$ is similar except
$\bar{P}_k=\ket{1}_1 \ket{1}_2\ket{1}_3...\ket{0}_k...\ket{1}_N$. D\"{u}r's state is a mixture of an $N$ qubit
GHZ state and collection of separable states and the loss of a qubit renders it separable. The
closest separable state has been given by Wei \cite{ref:Wei08} for $N\ge 4$
\begin{equation}
\sigma^{D\ddot{u}r}=
\frac{x}{2}\left(\matx{0}^{\otimes N}+\matx{1}^{\otimes N}\right)+\frac{1-x}{2N}\sum_{k=1}^N P_k +\bar{P}_k.
\end{equation}
This state is the same as the closest to the pure GHZ state mixed with the separable part of
$\rho^{D\ddot{u}r}$. Single qubit subsystems are maximally mixed states, $\Gamma=\Gamma_{sep}$ and all
geometric phase factors are trivial.

\subsection{$\Delta \gamma=\Delta \gamma_q$: State geometries described by entanglement}
\subsubsection{Mixtures of W and \={W} bar states}
The W states we consider here are the more traditional ones, in our notation $\ket{W}=\ket{N,1}$ and
$\ket{\bar{W}}=\ket{N,N-1}$. We consider $N$ qubit mixtures of these two states
\begin{equation}
\rho^{\bar{W}}=p\matx{W}+(1-p)\matx{\bar{W}}.
\end{equation}
Recently it has been shown that equal mixtures ($p=1/2$) of odd $N$
have no $N$ party classical correlations in the sense that all elements of the $N$ party correlation tensor
$\langle \sigma^1_{i_1} \sigma^2_{i_2} ... \sigma^N_{i_N} \rangle =0$. The indices $i_n$ can take the
values $x$, $y$ or $z$. However it was also shown that $\rho^{\bar{W}}$ has $N$ party entanglement meaning
there is no partitioning that can be written as a separable state \cite{ref:Kaszlikowski08}. The closest separable
state has been found by \cite{ref:Wei08} for a larger class, states of the form $\sum_k p_k \matx{N,k}$.
In general these $\sigma$ cannot be written down in closed form, also true for $\rho^{\bar{W}}$ for any $N$.
We can however write the closest separable states for $N=3,4$ in closed form. They are
\begin{equation}
\sigma^{\bar{W}}=\frac{1}{N^N}\sum_{l=0}^N \begin{pmatrix}
                                              N \\
                                              l \\
                                            \end{pmatrix} \alpha^l (N-\alpha)^{N-l}\matx{N,l}
\end{equation}
where $\alpha=p+(N-1)(1-p)$. The individual subsystems are given by
\begin{equation}
\rho_n=\frac{N-\alpha}{N}\matx{0}+\frac{\alpha}{N}\matx{1}.
\end{equation}
In the same way as we proceeded for pure W states in section~\ref{sec:previous} one can show
$\Gamma_{sep}=\sum_{n=1}^N \gamma_n^M$. Only
entanglement modifies the geometric phase. When $p=1/2$ all geometric phase factors are trivial.
Presumably for $N \ge 5$ when the closest
separable state becomes difficult to write down classical correlations become important in describing
$\Delta \gamma$.

\subsubsection{States resulting from tracing qubits out from $\ket{N,k}$}

One can also consider mixtures of symmetric states resulting from tracing qubits
out of $\ket{N,k}$. Provided
we consider states of $m\le k \le N-m$ qubits we have the entangled state
\begin{equation}
\rho^m=\sum_{l=0}^m \begin{pmatrix}
                      m \\
                      l \\
                    \end{pmatrix}\frac{\tiny \begin{pmatrix}
                                         N-m \\
                                         k-l \\
                                       \end{pmatrix}}
                    {\tiny \begin{pmatrix}
                       N \\
                       k \\
                     \end{pmatrix}}\matx{m,l}.
\end{equation}
The closest separable state has been found by \cite{ref:Vedral04}
\begin{equation}
\sigma^m=\frac{1}{N^m}\sum_{l=0}^m  \left(
                                      \begin{array}{c}
                                        m \\
                                        l \\
                                      \end{array}
                                    \right)
(N-k)^{m-l} k^l \matx{m,l}.
\end{equation}
The subsystems are still given by the same states as the W states in section~\ref{sec:previous} A similar
calculation as the one performed in that section shows $\Gamma_{sep}=\sum_{n=1}^m \gamma_n^M$. When $k=N/2$
all geometric phase factors are trivial.

\section{Properties responsible for $\Delta \gamma_q$}\label{sec:statecharacteristics}

What are the features of these states that put them either in the
$\Delta\gamma=\Delta\gamma_q$ or $\Delta\gamma=\Delta\gamma_c$ set? Using the geometric phase
we have been looking at geometrical properties of the projective Hilbert space invariant under
local $U(1)\times U(1)$ gauge transformations. In this section we look at some of
the properties of these states invariant under the action of local
$SU(2)$ gauge transformations, a higher symmetry group containing $U(1)\times U(1)$. They are also the same
transformations we have been making to obtain geometric phases. We actually look at
the invariants of the larger local special linear group
$SL(2,\mathbb{C})$ because several well known entanglement measures have this higher invariance as
well as invariance under $SU(2)$.

In this section we introduce and calculate a full set of $SL(2,\mathbb{C})$ invariants for
pure three qubit states. It is a full set in the sense that an arbitrary pure state of three qubits
can be determined up to local unitary transforms to a set of two possible states by the values of these invariants \cite{ref:Linden&Popescu98,ref:Sudbery01,ref:Acin01}.
All $SL(2,\mathbb{C})$ invariants we work with here are zero for the closest separable and
product states presented in this paper. This indicates these invariants may be useful
for discovering the features that make a difference to $\Delta \gamma_q$ but will not be useful
for identifying the properties responsible for $\Delta \gamma_c$.

\subsection{Local $SL(2,\mathbb{C})$ invariants}

\subsubsection{Bipartite entanglement $\tau_{ab}$}
To measure bipartite entanglement we will use
the square of the concurrence called the 2-tangle. It measures how entangled 2 qubits, $a$
and $b$ are and may be calculated from \cite{ref:Wootters98}
\begin{equation}
\tau_{ab}=\max\{0,\lambda_1-\lambda_2-\lambda_3-\lambda_4\}^2
\end{equation}
where $\lambda_i$ are the square roots of the eigenvalues of
$\rho_{ab}(\sigma_y\otimes\sigma_y)\rho_{ab}^*(\sigma_y\otimes\sigma_y)$ put in decreasing order.

All classes of states with $\Delta \gamma=\Delta\gamma_c$ have no bipartite entanglement suggesting it might be
responsible for $\Delta \gamma_q$. In general all $\Delta \gamma=\Delta\gamma_q$ states have
bipartite entanglement. However there are $\Delta \gamma=\Delta\gamma_q$ states with finite bipartite entanglement
and trivial geometric phase ($\Delta \gamma=0,\pi$) for example $\rho^{\bar{W}}$ for $N=3$ and $p=1/2$: The
2-tangle for each pair of qubits is $\tau=4/9[1-\sqrt{p(1-p)}]^2$. This suggests bipartite entanglement does
not uniquely prescribe the state space geometry due to entanglement.

\subsubsection{Tripartite entanglement $\tau_{abc}$}
For pure three qubit states Coffman \emph{et al.} introduced
the 3-tangle, a measure of how much entanglement there is in three way entanglement between the qubits.
The equation for pure state 3-tangle is given in \cite{ref:Coffman00}. To extend this notion to mixed three
qubit states we follow \cite{ref:Lohmayer06} and define the mixed state entanglement to be the average
pure state 3-tangle minimized over all possible decompositions of $\rho=\sum_i p_i \matx{\psi_i}$,
\begin{equation}
\tau_{abc}(\rho)=\min \sum_i p_i \tau_{abc}\left(\matx{\psi_i}\right).
\end{equation}
The expression for bipartite entanglement is defined analogously but has a known closed form.

There is only one state with non-zero 3-tangle, the $N=3$ GHZ state which has $\tau_{abc}=4\alpha^2\beta^2$.
We can exclude 3-tangle as the invariant responsible for $\Delta\gamma_q$.

\subsubsection{Twist $T(\tilde{a}\tilde{b}\tilde{c})$}
This was introduced in \cite{ref:Williamson&Wootters08}
as a quantity exhibiting $SL(2,\mathbb{C})$ invariance. They showed strong numerical evidence that
this invariant is a function of the Kempe invariant and therefore forms a complete set of local invariants
for pure three qubit states when accompanied with the three 2-tangles and 3-tangle. It is interesting as it arises
from an approach to generating $SL(2,\mathbb{C})$ and $SU(2)$ invariants inspired by lattice gauge theory. It
turns out this is the only non-trivial gauge field like invariant for pure states of three qubits and to calculate it you
construct a Wilson loop. The equation is
\begin{equation}
T(\tilde{a}\tilde{b}\tilde{c})=\frac{1}{4}\hbox{tr}\left[U(a,\tilde{c})U(c,\tilde{b})U(b,\tilde{a})\right].
\end{equation}
To obtain the unitaries $U(b,a)$ we take the $4\times 4$ correlation matrix
$S(b,a)_{ji}=\langle \sigma_i^a \otimes \sigma_j^b \rangle$, where $\sigma_i$ belong to the set of Pauli matrices $\{\mathbf{I},\sigma_1,\sigma_2,\sigma_3\}$ and polar decompose it into $U(b,a)=P^{-1}S(b,a)$.
$P$ is a positive semi-definite Hermitian matrix given by $\sqrt{S(b,a)S(b,a)^T}$. If $S$ is not of full rank
then $U$ is not unique and $T(\tilde{a}\tilde{b}\tilde{c})$ is undefined. The tildes denote a spin flip on that
particular qubit i.e. $U(b,\tilde{a})=U(b,a)\eta$ where $\eta=diag\{1,-1,-1,-1\}$.

The eigenvalues of $U(a,\tilde{c})U(c,\tilde{b})U(b,\tilde{a})$ are also $SL(2,\mathbb{C})$ invariants generally
being complex elements of $U(1)$. However for some states they are real and belong to $\mathbb{Z}_2$.
This occurs when $T(\tilde{a}\tilde{b}\tilde{c})=1$. We will term $T(\tilde{a}\tilde{b}\tilde{c})=1$ as
trivial twist in analogy with our terminology for the geometric phase.

All $\Delta \gamma=\Delta\gamma_c$ states have undefined twist making it a candidate for the invariant
responsible for $\Delta \gamma_q$. This is also supported by the fact $\Delta \gamma_q$ states have unique
and non-trivial twist when all geometric phases are non-trivial in all the
examples considered. To give some examples of values of twist, the state $\ket{3,1}$ has
$T(\tilde{a}\tilde{b}\tilde{c})=-0.41$ and $\rho_{\bar{W}}$, $N=3$ takes values between $0$ ($p=1/2$) and
$-0.41$ ($p=1$). This suggests that twist is the invariant responsible for $\Delta\gamma_q$. We have
found no counter example in the states considered.

\subsection{Discussion}

We have found that twist seems to be the most likely cause of the $\Delta\gamma_q$ correction to the geometric
phase. It is undefined for all the states with $\Delta \gamma=\Delta\gamma_c$ and becomes trivial or undefined
when the geometric phase becomes trivial ($\Delta\gamma=0,\pi$). Although we have not found a counter example
the results are not conclusive. The set of five $SL(2,\mathbb{C})$
invariants are only complete for pure three qubit states. For mixed states or states with higher numbers of qubits
further invariants must be added to completely describe the state up to local unitary equivalence. However, all
the $\Delta \gamma=\Delta \gamma_q$ states we considered were simple structures that first appear for pure states of three qubits or mixtures of such structures. Provided twist is the invariant giving rise to non-zero $\Delta\gamma_q$
one would also like to see the exact mechanism whereby the twist alters the geometric phase.

The other set of states that has $\Delta \gamma=\Delta\gamma_c$ are largely four qubit states. We have not been
able to identify which invariants are responsible for $\Delta \gamma_c$. To do this we could look at the
invariants between the entangled state, the closest separable state and the uncorrelated state. These should
be the same for the entangled and closest separable states but be different for the uncorrelated state. The
$SL(2,\mathbb{C})$ invariant set we have used here are not good for this purpose as they are zero or undefined
for all the unentangled states considered. Perhaps a set of $SU(2)$ invariants for 4 qubit states would be
useful for this purpose.

%\begin{tabular}{|l | l|}
%  \hline
%  % after \\: \hline or \cline{col1-col2} \cline{col3-col4} ...
%  $\ket{GHZ}$ & $-\alpha^2\log\alpha^2-\beta^2\log\beta^2$ \\
%  $\ket{N,k}$ & $-\log\left[{\tiny\begin{pmatrix} N \\ k \\ \end{pmatrix}} \left(\frac{N-k}{N}\right)^{N-k} \left(\frac{k}{N}\right)^k\right]$\\
%  $\ket{\phi^N}$ & $\frac{N}{2}$ for $N$ even $\frac{N-1}{2}$ for $N$ odd\\
%  $\rho^{Smolin}$ & 1 \\
%  $\rho^{D\ddot{u}r}$ & $x$ \\
%  $\rho^{\bar{W}}$ & $p\log\left[\frac{p N^{N-1}}{\alpha (N-\alpha)^{N-1}}\right]+
%  (1-p)\log\left[\frac{(1-p) N^{N-1}}{\alpha^{N-1} (N-\alpha)}\right]$\\
%  $\rho^m$ & $\sum_{l=0}^m \left[{\tiny \begin{pmatrix}m \\ l \\ \end{pmatrix}}\frac{\tiny\begin{pmatrix}N-m \\ k-l \\ \end{pmatrix}}
%  {\tiny\begin{pmatrix}N \\ k \\ \end{pmatrix}}\right]\log\left[\frac{\tiny\begin{pmatrix}N-m \\ k-l \\ \end{pmatrix}}
%  {\tiny\begin{pmatrix}N \\ k \\ \end{pmatrix}}\left(\frac{N}{N-k}\right)^{n-l}\left(\frac{N}{k}\right)^l \right]$\\
%  \hline
%\end{tabular}

\section{Conclusion}

In this paper we have shown that under the action of local unitaries correlations in
a state add corrections to the geometric phase in addition to the local phases obtained
by its subsystems. We showed that correlations did not change the other phase in quantum
mechanics, dynamical phase.

Of the two types of correlation in quantum mechanics, entanglement and
classical correlations, we showed that this correction to the geometric phase could be
described entirely by classical correlations for GHZ states, cluster states and two examples
of bound entangled state. In contrast we found that this correction could be completely
described by entanglement for W states and mixtures of W states.

We investigated what properties of the state may be responsible for the entanglement correction
to the geometric phase using local invariants to probe structures of the states more finely.
We found one possible candidate for the entanglement correction, this was a quantity called twist, also
geometrical in construction and of a possible gauge field like interpretation.

Regarding possible future work, this study restricted the subsystem evolutions to be cyclic, that is each
subsystem came back to itself after some arbitrary unitary evolution. It would be interesting to consider
the non-cyclic cases. We also restricted the symmetries of the subsystem paths in the projective Hilbert
space to be abelian ($U(1)\times U(1)$) even when the subsystems became maximally mixed. When the subsystems
become maximally mixed the symmetry becomes elevated to $U(2)$ and one may obtain a non-abelian geometric phase.
This may also be interesting to investigate further. It would also be nice to see the exact mechanism that
produces these corrections to the geometric phase and identify possible properties that result in the
classical correlation correction to the geometric phase.

MSW acknowledges partial funding from EPSRC and QIP IRC www.qipirc.org (GR/S82176/01). We
thank Michal Hajdu\v{s}ek for communicating his results on cluster states.

\bibliography{masterbib}
\bibliographystyle{unsrt}

\end{document}